\documentclass[twoside,leqno,twocolumn]{article}

\usepackage[letterpaper]{geometry}

\usepackage{ltexpprt}

\usepackage{hyperref}

\usepackage{cleveref}

\usepackage{svg}

\usepackage{subcaption}

\usepackage{algorithm}
\usepackage{algpseudocode}

\usepackage{listings}
\usepackage{xcolor}

\usepackage{booktabs}

\usepackage{rotating}

\usepackage{multirow}

\usepackage{amssymb}

\newtheorem{definition}{Definition}
\newtheorem{prooof}{Proof}

\begin{document}

\newcommand\relatedversion{}
\renewcommand\relatedversion{\thanks{The full version of the paper can be accessed at \anon{\protect\url{redacted-url}}{\protect\url{https://arxiv.org/abs/1902.09310}}}} 

\title{\Large Exploring polyhedral mesh  generation from Delaunay tetrahedral meshes }
 { 
    \author{Sergio Salinas\thanks{ssalinas@dcc.uchile.cl, DCC, Universidad de Chile, Chile.}
    \and Magdalena Alvarez\thanks{DCC, Universidad de Chile, Chile.}
    \and  Nancy Hitschfeld\thanks{nancy@dcc.uchile.cl, DCC, Universidad de Chile, Chile.}}    
}

\date{}

\maketitle


\fancyfoot[R]{\scriptsize{Copyright \textcopyright\ 2025 by SIAM\\
Unauthorized reproduction of this article is prohibited}}





\begin{abstract} \small\baselineskip=9pt 
In this paper we propose a new algorithm to generate polyhedral meshes based on the concept of terminal-face regions. Terminal face-regions are built from the union of tetrahedra by a face that fulfill some joining criterion. For this, we introduce first the concept of terminal-face region and how it can be seen as a natural extension of 2D terminal-edge regions ~\cite{Salinas-Fernandez2022}. Next, we include the related work in 3D polyhedral mesh generation and then the algorithm we have designed and implemented to generate polyhedral meshes. Finally, we present statistics about the kind of generated polyhedral meshes, and also a theoretical comparison with Voronoi meshes.
\end{abstract}

\section{Introduction}

Going from flat (2D) meshes to 3D shapes (volumetric meshes) is a big step in computer science and engineering simulations. Volumetric meshing divides a 3D object into tiny pieces to exactly represent its volume, which is important for running simulations. Mesh generation methods are important in science, engineering, and computer graphics. Tetrahedral meshes~\cite{Yerry_Shephard,Shephard_90,Pascal96,tetgenxd,TetWild2018} and hexahedral meshes~\cite{Pietroni23,Gargallo-PeiroR15} are chosen based on the problem and the used numerical method. In addition, there are approaches to generate mixed element meshes~\cite{YamakawaS11,Hitschfeld2004} and polygonal and polyhedral meshes~\cite{Abdelkader2020,GarimellaKB13,PolyMesher2012,YanWLL10,EbeidaM11,SiegerAB10}, with polygonal/polyhedral cells as Voronoi regions too. Mesh generation algorithms can be divided into direct and indirect algorithms~\cite{owen1998survey}, with the latter having the advantage of using robust open source for initial meshes, like Tetgen~\cite{tetgenxd}, Tetwild~\cite{TetWild2018}, and CGAL~\cite{cgal:rty-m3-24a}.

With respect to the numerical methods, the Finite Element Method (FEM) has been expanded to include general polyhedral cells, and a new method called the Virtual Element Method (VEM) has been created for 2D and 3D problems. Research is now ongoing to improve and evaluate the VEM in different applications~\cite{Attene2019,SBMS2022,SBMS2022a,SorgenteSirvey}. VEM has been already applied in various fields, including solid mechanics~\cite{da2015virtual,Wriggers2017,Wriggers2017a}, fluid dynamics~\cite{caceres2015mixed}, and modeling of brittle crack propagation~\cite{HUSSEIN201915}.

In water flow simulations of large areas, unstructured grids are needed. A tool called MODFLOW-USG~\cite{Panday2013} can handle different mesh setups. It uses a Control Volume Finite Difference (CVFD) approach, requiring a line connecting the centers of two adjacent cells to intersect the mutual boundary perpendicularly. This mirrors earlier efforts in simulating semiconductor devices with polyhedral Delaunay meshes with Voronoi regions as control volumes~\cite{Hitschfeld92,Hitschfeld2004}.

Due to the requirements to count with  polyhedral mesh generators that allow researchers to evaluate how general polyhedral cells can be and still give accurate simulation results, we present a first attempt to study the kind of polyhedra that  can be generated by joining tetrahedra from a tetrahedral mesh. We present an algorithm and  describe the properties of the built  polyhedral cells. The preliminary experimental evaluation shows that the generated polyhedra are very far from convex cells and need to be further study too see if these kind of cells are useful and provide some advantages in the context of some specific simulation in comparison to  Voronoi cells. A mesh composed of proper non-convex polyhedral cells should require less cells and points to model complex geometries than meshes composed of convex cells. Currently, together with a good point distribution, the strongest requirement for the VEM cells seems to be that  polyhedron kernels should not be  empty~\cite{SorgenteSirvey}. 


\section{Basic concepts}
\label{chapt3Dsec:basic_concepts}

In this section we introduce the concepts needed to understand the proposed algorithm. Several concepts  are  lemmas and theorems presented in~\cite{Rivara97, Alonso2018DelaunayBA, Salinas-Fernandez2022}, but others are new concepts and definitions to handle  tetrahedral meshes.

We want to extend the concept of terminal-edge region, defined in ~\cite{Salinas-Fernandez2022}, to faces in 3D, and convert them to polyhedrons to generate volume meshes. One of the problem of doing this extension, is what is ``the largest face of a tetrahedron'', in such a way that we can define a path of tetrahedra. There are several criteria to define which face of a tetrahedron is the largest or the smallest, we are going to call them ``Joining criteria''. Given a tetrahedral mesh $\tau = (V,E,F)$, we can define:

\begin{definition}\label{def:JoinCriterion}
    \textbf{Joining criterion}  For any tetrahedron $t_i$, the Joining criterion is a metric used to rank the faces of $t_i$ ordered from the largest to the smallest. There is only one largest-face, if two faces have the same size, then one of them is chosen as the largest arbitrarily.   
\end{definition}

A joining criterion can be for example the area of the face, then we can rank the faces of the tetrahedron according to which tetrahedron has the largest and the smallest face area. With this already defined, we can extend the concept of terminal edge and longest-edge propagation path~\cite{Rivara97} to faces in a tetrahedral mesh $\tau$ with a Joining criterion $J$:

\begin{definition}
    \textbf{Terminal-face} A face is a terminal-face $f_i$ if two adjacent tetrahedrons $t_a, t_b$ to $f_i$ share their respective (common) largest-face, according to $J$. This means, that $f_i$ is the largest-face of both tetrahedrons that share $f_i$. If $t_b = \emptyset$, then $f_i$ is called border terminal-face. \\
\end{definition}

\begin{definition}\label{def:Lfpp}
    \textbf{Largest-face propagation path (Lfpp)}  For any tetrahedron $t_0$ of any tetrahedralization $\tau$, the Largest-Face Propagation Path of $t_0$ ($Lfpp(t_0)$) is the ordered list of all the tetrahedrons $t_0, t_1, t_2,  ..., t_{n-1}$,  such that $t_i$ is the neighbor tetrahedron of $t_{i-1}$ by the largest face of $t_{i-1}$, for $i = 1,2, ..., n-1$.  \\
\end{definition}

For a terminal-face $f_i \in \tau$, there are various Lfpps that ends in the same $f_i$, it is the case even when $f_i$ is a border face, we are going to call the set of Lfpp with the same terminal face a terminal-face region.

\begin{definition}
        \textbf{Terminal-face region} A {\em terminal-face region} $R$ is a region formed by the union of all tetrahedrons $t_i$ such that Lfpp($t_i$) has the same terminal-face. \\
\end{definition}

\begin{figure}
	\centering
	\includegraphics[width=1\linewidth]{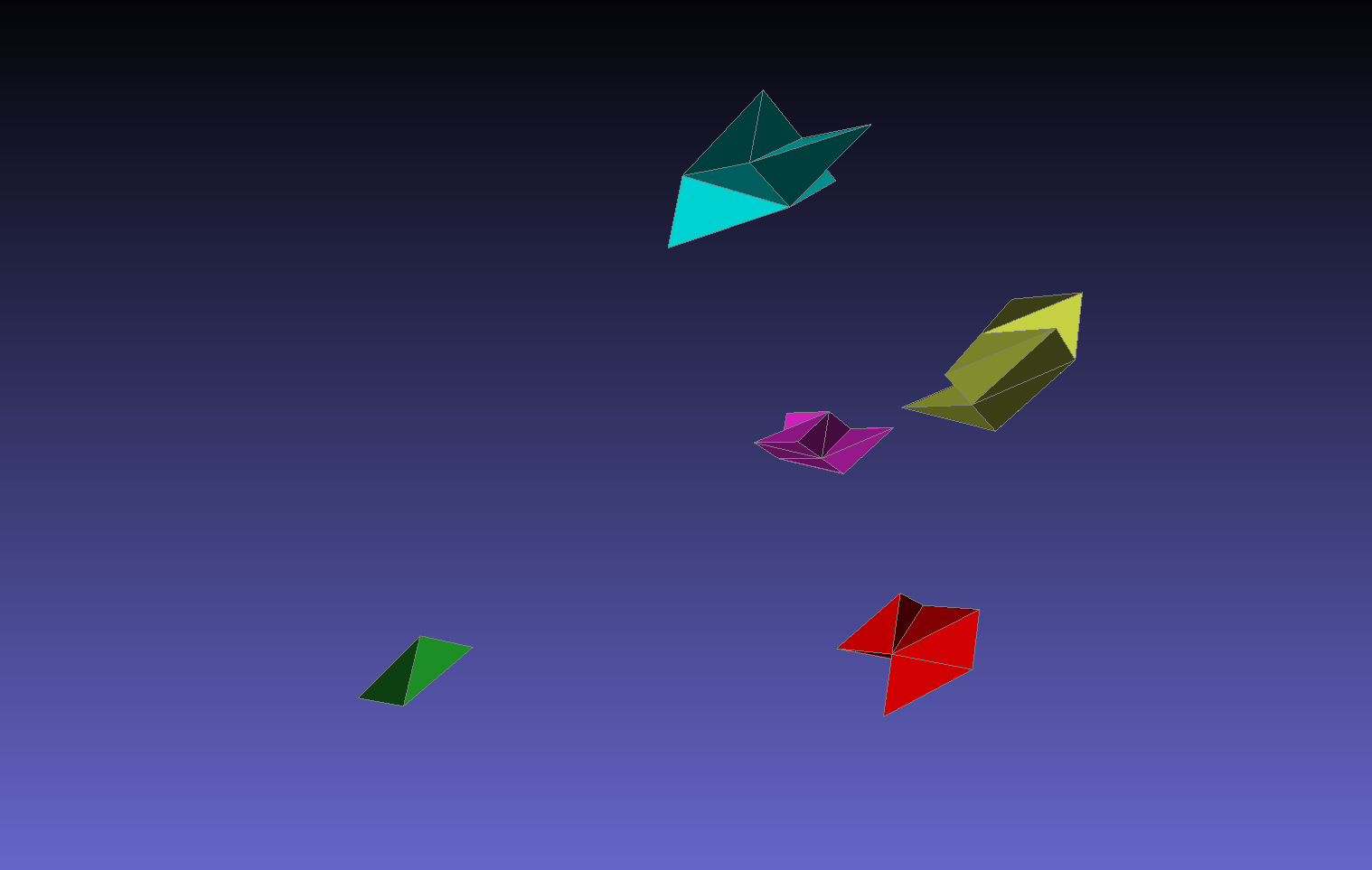}
	\caption{Example of five terminal-faces region generated using the incircle joining criterion.}
	\label{fig:terminalfaces}
\end{figure}

\begin{figure}[]
    \centering
    \begin{subfigure}[b]{0.45\textwidth} 
        \centering
        \includegraphics[width=\textwidth]{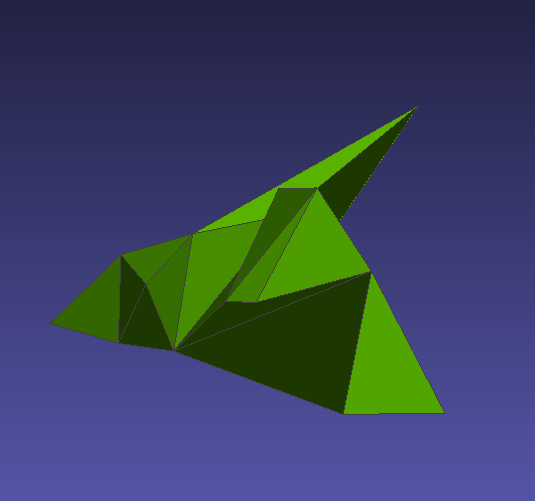} 
        \caption{Polyhedron generated with the incircle criterion}
        \label{fig:incirclepolyhedra}
    \end{subfigure}
    \hfill 
    \begin{subfigure}[b]{0.45\textwidth} 
        \centering
        \includegraphics[width=\textwidth]{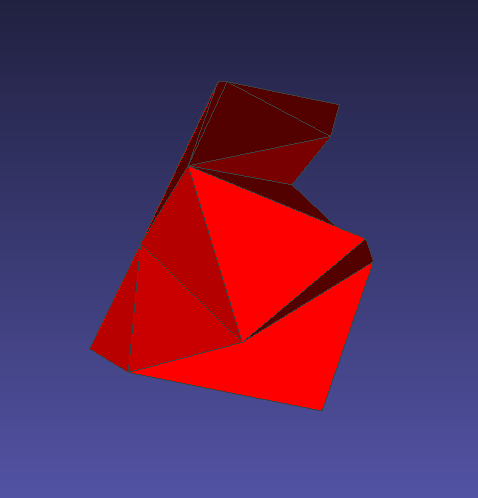} 
        \caption{Polyhedron generated with the max area criterion}
        \label{fig:maxareapolyhedra}
    \end{subfigure}
    \caption{Comparison of two polyhedra with a greater number of faces, derived from the same tetrahedralization, generated using two different joining criteria.}
    \label{figs:polyhedra_joinin_cretia}
\end{figure}

Figure~\ref{fig:terminalfaces} illustrates examples of terminal-face regions generated using the incircle criterion. Meanwhile, Figure~\ref{figs:polyhedra_joinin_cretia} displays the largest polyhedron, derived from the same tetrahedralization, created using two different joining criteria. With the concepts defined above, we have a proper extension of the terminal-edge region to faces in tetrahedral meshes. 

In order to accelerate the generation of terminal-face regions we develop a classification system for each face of $\tau$. For each face $f_i \in \tau$, a face can be a frontier-face or internal-face.

\begin{definition}
    \textbf{Frontier-face} A frontier-face $f_i$ is a face that is shared by two tetrahedron $t_1, t_2$, each one belonging to a different terminal-face region, that means that $f_i$ is not the largest-face of neither $t_1$ nor $t_2$. If $t_2 = \emptyset$ then $f_i$ is a frontier-face even if $ef_i$ is a border terminal-face. \\
\end{definition}

\begin{definition}
    \textbf{Internal-face} A internal-face $f_i$ is a face that is shared by two tetrahedrons $t_1, t_2$, each one belonging to a same terminal-face region. In other words, $f_i$ is an internal-face if $f_i$ is neither a terminal-face nor frontier-face.\\
\end{definition}

For the context of this work, we only need the frontier-faces, as those faces will be faces of the polyhedral mesh at the end of the algorithm, internal-faces will be removed during the process of joining tetrahedrons. 
One property that terminal-face regions must obey, for their use as polyhedrons in a polyhedral mesh, is that they must not overlap.

\begin{lemma}
Let \( \tau \) be a tetrahedral mesh of any set of points \( P \) with a Joining criterion $J$. Then the set of terminal-face regions in \( \tau \) do not overlap.
\end{lemma}

\begin{prooof}  \emph{By contradiction.} 
Let us assume the tetrahedron $t_k$ belongs to two terminal-face region $R_i, R_j$, each one with the terminal-faces $f_i, f_j$, respectively. Since $t_k$ has only one largest-face, Lfpp($t_k$) has to end in either $f_i$ or $f_j$, but by definition of the Joining criterion, $t_k$ only have one largest-face, then, $t_k$ only can have one Lfpp, thus Lfpp($t_k$) have to end either $f_i$ or $f_j$ but not both. 

Hence,  a tetrahedron $t_k$ can not belong to two terminal-face regions $R_i, R_j$, and therefore there are not overlapping terminal-face regions in $\tau$. $\blacksquare$

\end{prooof}

As in 2D, terminal-edge regions can contain frontier-edges in its interior, in 3D, terminal-face regions can contain frontier-faces in their interior. We will refer to this kind of frontier faces as barrier-faces.

\begin{definition}{\textbf{Barrier-face}}\label{d:barrier-edge}
    Given a terminal-face region $R_i$, any frontier-face $f \in R_i$ that is not part of the boundary $\delta  R_i$ is called a barrier-face. \\
\end{definition}

\begin{definition}{\textbf{Barrier-edge tip}}\label{d:barrier-edge}
    A barrier tip in a terminal-face region $R_i$ is an edge incident to only one frontier-face (particularity a barrier-face), and the rest of the faces are internal-faces.
\end{definition}

\section{Algorithm}
\label{chapt3Dsec:polylla_face}

In this section, we present our algorithm to generate polyhedral meshes. 
 The algorithm takes as input a tetrahedral mesh $\tau = (V, E, F)$, a Joining criterion $J$, and return a polyhedral mesh $\tau' = (V, E', F')$. The algorithm has 3 main phases: Label phase, Traversal phase and Repair phase.

For the understanding of the algorithm, we are going to assume that we have a data structure with all geometrical information of the tetrahedral mesh. For the output we use a list of polyhedrons, and each polyhedron is represented as a list of faces.

\subsection{Label phase}

The first step to generate the new polyhedral mesh is to define whose faces are going to be faces of the output mesh (i.e. the frontier-faces), and choose one tetrahedron $t_i$ per terminal-face region $R_i$, to be used in the Traversal phase to generate the new polyhedron, we are going to name to $t_i$ as seed tetrahedron, and it is a tetrahedron adjacent to a terminal-face. For this reason, as equal as the 2D algorithm, we first with the Label phase. 

The Label phase receives $\tau = (V, E, F)$ as input, return two auxiliary arrays with information of the mesh:

 \begin{itemize}
    \item \texttt{Seed array}, of equal size to the number of tetrahedrons, with the indices of all the seed tetrahedrons in $\tau$.
    \item \texttt{Frontier-face bitvectors}, of equal size to the number of faces $|F|$, in where each element $i$ is set as true if the face $f_i$ is a frontier-face, and false is not. %
 \end{itemize}

Those auxiliary arrays are equivalent to the auxiliary data structure used in the 2D algorithm for the polygonal mesh generation. Those arrays will be use in the Traversal phase and the Repair phase.

For this phase, we first define a join criterion $J$, and each face $f_i \in \tau$ is labeled according if this accomplishes $J$. Examples of join criteria for a tetrahedron $t_i \in  \tau$ are:

\begin{itemize}
    \item Area criterion: The largest face of $t_i$ is the face with maximum area of $t_i$ and the smallest face is the face with minimum area of $t_i$.
    \item In-circle radius criterion: The largest face of $t_i$ is the face with maximum in-circle radius of $t_i$ and the smallest face is the face with minimum in-circle radius of $t_i$.

\end{itemize}

Depending on the joining criterion, the properties of the polyhedral mesh will change, in the experiments section we are going to show examples of meshes with those two criteria.

This phase is shown in Algorithm~\ref{algo:3dfacelabelphase}. The algorithm first calculates the faces that accomplish the joining criterion $J$ of each tetrahedron $t_i \in \tau$, it is shown in line~\ref{3Dfacealgolabel:tetraIteration} --~\ref{3Dfacealgolabel:tetraIteration2}, for each $t_i$, the algorithm calculates the largest face $f_j$ according to $J$, and stores it in the array \texttt{largest\_array}, this array has size equal to the number of tetrahedrons in the mesh, an store the index $j$ of the face $f_j$ in the position  $i$, corresponding to the position of $t_i$ in the array. For example, for a tetrahedron $t_i$ with the area criterion, the algorithm compares the four faces of $t_i$ and stores the index of the largest face in \texttt{largest\_array}.

Afterwards, the algorithm labels the seed tetrahedrons (lines~\ref{3dfacealgolabel:largest_teration} --~\ref{3dfacealgolabel:largest_teration2}), those are the tetrahedrons adjacent to a terminal-face, and that are used in the traversal phase to generate the polyhedrons. For each face $f_i \in \tau$, the algorithm gets both tetrahedons $t_i, t_j$ and chooses one them as a seed-edge an stores it in the \texttt{seed\_array}. If $f_i$ is a border face, then it chooses only a adjacent tetrahedron as seed.

\begin{figure}[h]
    \centering
    \includegraphics[width=0.9\linewidth]{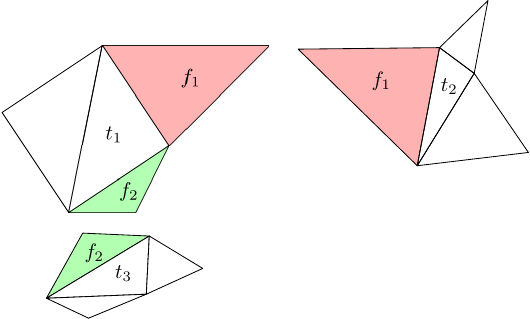}
    \caption{Example of labeling with adjacent $3$ tetrahedrons, Red faces are the terminal-faces, and green faces are frontier-faces. Tetrahedron $t_1$ is connected to $t_2$ by face $f_1$, and tetrahedron $t_1$ is connected to $t_3$ by face $f_2$. According to the Joining criterion of the largest area, $f_1$ is the largest face of $t_1$ and $t_2$, meaning that $f_1$ is a terminal-face, thus $t_1$ is chosen a seed tetrahedron. $f_2$ is not the largest face of $t_1$ and $t_3$, thus $f_2$ is label as a frontier-face.}
    \label{fig:3dlabelphase}
\end{figure}

Finally, the algorithm labels the frontier-faces (lines \ref{3dfacealgolabel:seediteration} --~\ref{3dfacealgolabel:seediteration2}), those are the faces of the final mesh $\tau'$. For each face $f_i \in \tau$, the algorithm gets the both tetrahedrons $t_i, t_j$, that shares $f_i$, if $f_i$ is not the largest face of neither $t_i$ nor $t_j$, or if $f_i$ is a border face, then $f_i$ is labeled as frontier-edge (sets as true) in the \texttt{frontier\_Bitvector}. This process is exemplify in Figure~\ref{fig:3dlabelphase}.

With the tetrahedrons and faces already labeled, the algorithm continues to the Traversal phase.

\begin{algorithm}[H]
    
    \caption{Label phase}\label{algo:3dfacelabelphase}
    \begin{algorithmic}[1]
    \Require Tetrahedral mesh $\tau$
    \Ensure Bitvectors {\tt frontier-face} and {\tt max-face}, and vector {\tt seed-list}

    \ForAll{tetrahedron $t_i$ in  {\tt tetrahedron\_array}} \label{3Dfacealgolabel:tetraIteration} \Comment{Calculate largest face}
        \State Calculate the join criteria $J$ of all faces $f_1, f_2, f_3, f_4 \in t_i$
        \State $f_{max}$ $\leftarrow$ $max(f_1, f_2, f_3, f_4)$ 
        \State Append $f_{max}$ to \texttt{largest\_face\_array }
    \EndFor \label{3Dfacealgolabel:tetraIteration2}
    
    \ForAll{face $f_i$ in {\tt face\_array}} \label{3dfacealgolabel:largest_teration}
    \Comment{Label seed tetrahedron}
        \State $t_i, t_j$ $\leftarrow$ adjacent tetrahedrons to $f_i$
        \If{$t_i = \emptyset$ and  $f_i$ is the largest-face of $t_i$} 
            \State \texttt{seed\_face\_array}[$f_i$] = $t_j$
        \ElsIf{$f_i$ is the largest-face of both $t_i$ and $t_j$ }
            \State \texttt{seed\_face\_array}[$f_i$] = $t_i$
        \EndIf
    \EndFor \label{3dfacealgolabel:largest_teration2}
    \ForAll{face $f_i$ in {\tt face\_array}} \label{3dfacealgolabel:seediteration}
    \Comment{Label frontier-faces}
        \State $t_1, t_2$ $\leftarrow$ adjacent tetrahedrons to $f_i$
        \If{$t_1 = \emptyset$ or $t_2 = \emptyset$} 
            \State \texttt{frontier\_Bitvector}[$f_i$]  = True
        \ElsIf{\texttt{largest\_array}[$t_i$] is not $f_i$  and \texttt{largest\_array}[$t_j$] is not $f_i$}
            \State \texttt{frontier\_Bitvector}[$f_i$]  = True
        \EndIf
    \EndFor \label{3dfacealgolabel:seediteration2}
\end{algorithmic}
\end{algorithm}

\subsection{Traversal phase}

\begin{figure}
\centering
    \includegraphics[width=0.8\linewidth]{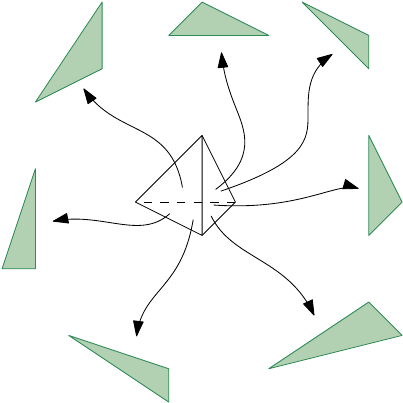}
    \caption{Visualization of the DFS, the tetrahedron  at the center is a seed tetrahedron, the DFS travels inside a terminal-face region until to find a frontier-face (green faces)  and stores it as a face of the new polyhedron.}
    \label{fig:3Dtraversal}
  \end{figure}

In this phase the algorithm converts terminal-face regions into polyhedrons. To do this, for each tetrahedron $t_i \in$ \texttt{seed\_array}, the algorithm defines a polyhedron $P$ and calls to the depth first search (DFS) algorithm~\cite{Levitin} shown in Algorithm~\ref{algo:3dtravelphase}. In this DFS, the algorithm travels inside the terminal-face region using the faces of the seed tetrahedron $t_i$. For each tetrahedron $t_j$ adjacent to $t_i$ by its face, the algorithm checks if $t_j$ contains a frontier-face $f_i$, if it is true, then $f_i$ is stored in $P$, as part of the polyhedron, and if its is not the case, then $f_i$ is a internal-face. Consequently, the DFS travel to the neighbors of $t_j$ looking for others frontier-faces. This process is shown in Figure~\ref{fig:3Dtraversal}.

\begin{algorithm}[H]
    \caption{Depth First Search for polyhedron construction}\label{algo:3dtravelphase}
    \begin{algorithmic}[1]
    \Require Seed edge $e$ of a terminal-edge region
    \Ensure Arbitrary shape polyhedron $P$

    \State $P \leftarrow$ $\emptyset$

    \Procedure{DepthFirstSearch}{Seed Tetrahedron $t_i$, Polyhedron $P$}
    \State Mark $t_i$ as visited
    \ForAll{neighbor Tetrahedron $t_j \in t_i$ }
        \If{common face $f_i$ of $t_i, t_j$ is a frontier-face}
            \State Add $f_i$ to polyhedron $P$
        \Else
            \If{$t_i$ has not been visited yet}
                \State DepthFirstSearch($t_j$, $P$)
            \EndIf
        \EndIf
    \EndFor
    \EndProcedure
    \end{algorithmic}
\end{algorithm}

The following Lemma demonstrates that this DFS will only travel inside a terminal-face region.

\begin{lemma} Depth First Search algorithm presented in algorithm~\ref{algo:3dtravelphase} only travels inside one terminal-face region $R_i$
\end{lemma}

\textbf{Proof} Suppose that exist a terminal-face region $R_j$, adjacent to $R_i$, both connected for the frontier-face $f$, which is part of the tetrahedrons $t_i \in R_i$ and $t_j \in R_j$. If during the DFS the algorithm passes from to $R_i$ to $R_j$ means that $f$ is not labeled as a frontier-edge, but by definition, if $f$ is the separation between $R_i$ and $R_i$, then $f$ must be a frontier-face. Thus for contradiction the DFS only travels inside one terminal-face region. $\square$

Notice that can occur the case that the terminal-face contains barrier-edge tips, thus, for each $P$ generated from the DFS, the algorithm checks them. The algorithm counts the number of repeated faces in $P$, if there are repeated faces, it means that a face was stored two times during the DFS, indicating a barrier-face. In such cases, $P$ is sent to the Repair phase.

\subsection{Repair phase}

For any not simple polyhedron $P_i$, the algorithm uses the barrier tips to split a polyhedron in two. A barrier tips is an edge $e_i \in P_i$ that is adjacent to only one frontier-face of $P_i$, and the rest of faces adjacent to $e_i$ are internal-faces. The first question to answer is how to know if edge $e_i$ is a barrier tip.

\begin{theorem}\label{teo:repairphase}
    Given a terminal-edge region $R_i$, with $F_f$ the set of frontier-faces of $R_i$, an edge $e$ belonging to a barrier-face of $R_i$, and the set $F_e$ of faces incident to $e$., $e$ is a barrier tip if $|F_f| - |F_e \cap F_f| = |F_f| - 1$. 
\end{theorem}

\begin{prooof} By contradiction. Assume that $e$ is a barrier tip and $|F_e \cap F_f| \not= 1$:
    
\begin{itemize}
    \item $|F_e \cap F_f|$ is more than $1$: then there exist more than one barrier-face incident to $e$, but by Definition~\ref{d:barrier-edge}, $e$ only can be incident to a barrier-face, so $e$ is not a barrier tip. 
    \item $|F_e \cap F_f|$ less than $1$: then there is no barrier-faces (neither frontier-faces) incident to $e$, but by Definition~\ref{d:barrier-edge}, $e$ have to be incident to a frontier-face. So $e$ is not a barrier tip. 
\end{itemize}
    
Therefore, the cardinal of $|F_e \cap F_f|$ is equal to $1$, and $e$ must fulfill that $|F_f| - |F_e \cap F_f| = |F_f| - 1$. $\blacksquare$

\end{prooof}

Using Theorem~\ref{teo:repairphase} we define the Algorithm~\ref{algo:barrierfacedetection} to get a list $B_p$ with all the barrier tips of polyhedron $P$. The algorithm takes all edges $e_i$ of the barrier-faces in $P$, and iterates over all them (line~\ref{algo:barrierface}), to check if they are a barrier tip. The check is done by calculating (line~\ref{algo:barrierface2}),  the formula of above, if it is true, then there is only a barrier-face adjacent to $e_i$, thus $e_i$ is barrier tip.

\begin{algorithm}[H]
    \caption{Barrier-face Detection}\label{algo:barrierfacedetection}
    \begin{algorithmic}[1]
    \Require Polyhedron $P_i \in tau'$, $F_b$ barrier-faces $\in P_i$
    \Ensure List of barrier tips $B$
    
    \State $B \leftarrow \emptyset$ \Comment{List of barrier tips} 
    \ForAll{edges $e_i \in F_b$} \Comment{For all the edges of the barrier-faces of $P$} \label{algo:barrierface} 
        \State $F_e \leftarrow$ List of all faces incident to $e_i$
        \State $|F_f| \leftarrow$ number of frontier-face in $P$
        \State $|F_e \cap F_f| \leftarrow$ number of faces in $F_e$ that are also in $F_f$
        
        \If{$|F_f| - |F_e \cap F_f| = |F_f| - 1$} \label{algo:barrierface2} 
            \State $B \leftarrow B \cup \{e\}$ \Comment{$e$ is a barrier tip}
        \EndIf
    \EndFor
    \State \textbf{return} List of barrier tips $B_p$
    \end{algorithmic}
\end{algorithm}

Once the algorithm has computed the set of barrier tips $B$, we can use them to split the polyhedron $P$. This split consists of converting internal-faces $f_i$ to frontier-faces, and using the two tetrahedron adjacent to $f_i$ as seeds to repeat the traversal phase.

This repair phase is shown in Algorithm~\ref{algo3D:reparationphase}.  The algorithm first defines a \texttt{subseed list} $L_p$ to store the seed tetrahedrons that will be used as seeds to generate the new polyhedra. And the \texttt{usage bitarray} $A$ that is used as a flag to check if a seed tetrahedron has been used during the creation of a new polyhedron, thus the algorithm can avoid creating duplicate polyhedra.

Then, in line~\ref{algorepa3D:foreachbet}, the algorithm iterates over all the barrier tips $b_i \in B$. For each $b_i$, the algorithm selects the barrier-face $f_i$ incident to $b_i$, circles around the internal faces of $b_i$, and stores them in order of appearance in a sublist $l$. The middle internal face $f_m$ of $l$ is calculated. $f_m$ is converted to a frontier-face by setting \texttt{frontier\_bitvector}[$f_m$] = \texttt{True}. The two tetrahedron $t_1$  and $t_2$ adjacent to $f_m$ are stored in the list $L_p$ to be used as seed tetrahedra, and they are also marked as \texttt{True} in the \texttt{usage\ bitarray} $A$.

Later, in line~\ref{algorepa3D:foreachseedtriangle}, the algorithm constructs the polyhedron. For each tetrahedron $t_i \in L_p$, the algorithm checks if $t_i$ has been used during the generation of a tetrahedron. If this is not the case, then the algorithm proceeds to generate a new polyhedron $P'$ by calling the traversal phase shown in Algorithm~\ref{algo:3dtravelphase}. However, for each tetrahedron $t_j$ visited in the traversal phase, A[$f_m$] is set to \texttt{False} to avoid using $t_j$ to generate the same polyhedron $P'$ again. This process is repeated until there are no more seed tetrahedron in $L_p$, at which point all the new polyhedrons are simple polyhedrons, and are added to $\tau'$.

\begin{algorithm}[H]
    
    \caption{Non-simple polyhedron reparation}\label{algo3D:reparationphase}
    \begin{algorithmic}[1]
    \Require Non-simple polyhedron P, list of barrier-faces tip $B$
    \Ensure Set of simple polyhedron $S$ 
    \State \texttt{subseed list}  as $L_p$ and \texttt{usage bitarray} as $A$ \label{algorepa3D:inithash}
    \State $S$ $\leftarrow$ $\emptyset$ 
    \ForAll{barrier-faces tip $b_i$ in $B$} \label{algorepa3D:foreachbet}
        \State $f_i$ $\leftarrow$ Barrier-face that contains $b_i$
        \State Calculate the valence of $b_i$
        \State $f_m$ $\leftarrow$ middle-face incident to $b_i$        
        
        \State Label $f_m$ as frontier-edge \label{algorepa3D:labelmidedge}
        \State Save tetrahedrons $t_1$ and $t_2$ adjacent to $f_m$ in $L_p$ \label{algorepa3D:savebothhalfedges}
        \State $A[t_1] \leftarrow $ \texttt{True}, $A[t_2] \leftarrow $ \texttt{True}  \label{algorepa3D:savetobivector}
        
    \EndFor

    \ForAll{tetrahedrons $t_i$ in $L_p$} \label{algorepa3D:foreachseedtriangle}
        \If{$A[t_i]$ is \texttt{True}} \label{algorepa3D:true}
            \State $A[t_i] \leftarrow $ \texttt{False}
            \State Generate new polyhedron $P'$ starting from $t_i$  \label{algorepa3D:generationpoly} repeating the Traversal phase. 
        \State Set as \texttt{False} all indices of tetrahedron in $A$ used to generate $P'$ \label{algorepa3D:removeseeds}
        \EndIf  
        
    \State $S$ $\leftarrow$ $S$ $\cup$ $P'$ 
    \EndFor 
    \State \Return $S$

    \end{algorithmic}
\end{algorithm}

Finally, $\tau'$ is a mesh composed of simple polyhedra. Notice that a more simple strategy to generate simple polyhedra is just to eliminate the barrier faces because they are internal faces and do not represent geometrical aspects to be respected from the input domain. 

\section{Experiments}\label{sec:experiments}

In this section we present the preliminary results. To study the geometrical properties of the generated polyhedrons we developed a first prototype in Python. This prototype uses 4 array of structures to store the information of vertices, edges, faces and tetrahedrons of a tetrahedral mesh. 

The experiments consist in testing the algorithm with $4$ different kinds of tetrahedral meshes. The tetrahedral meshes  are the following:

\begin{itemize}
    \item Random meshes: We generate random points using a uniform distribution inside a cube, with method \texttt{random.rand} of python's library Numpy, and call to the method \texttt{scipy.spatial.Delaunay} in the python's library SciPy to generate Delaunay tetrahedral mesh. 
    \item Poisson meshes: We generate random points using a Poisson distribution inside a cube, using the method \texttt{scipy.stats.qmc.PoissonDisk} of Scipy. and call to the method ``Delaunay'' in the python's library SciPy to generate Delaunay tetrahedral mesh.
    \item Quality Delaunay meshes: We call to Tetgen with the python interface of PyVista with the parameters \texttt{pqa[number]}, where \texttt{pq} generated a  boundary conforming quality tetrahedral, with a maximum radius edge ratio bound as $2.0$ and a minimum dihedral angle bound $0$ degrees, respectively. And \texttt{a[number]} define a maximum tetrahedron volume constraint to the mesh. We define different mesh sizes by changing the value of the volume constraint \texttt{number}, the used values were $0.0008$, $0.00038$,$0.0000695$, $0.000033$, and $0.000006279$, for the meshes with $496$, $992$, $4993$, $9988$ and $49999$ vertices respectively.
    \item Grid meshes: We create equidistant points in a cube, and create a structure grid using the PyVista library, this grid is converted to a tetrahedral mesh using the method ``delaunay\_3d'' of PyVista. 
\end{itemize}

Notice that Random, Poisson and Grid meshes refers to a point distribution, but Quality meshes is a kind of mesh obtained from a mesh refinement. We used Tetgen because it is a popular mesh generator, and because the quality criteria of the tetrahedrons seems to be good intermediate between the quality of the tetrahedrons of the Poisson meshes and Grid meshes.

\begin{figure*}
    \centering
    \begin{subfigure}{0.45\textwidth}
    \centering
    \includegraphics[width=\textwidth]{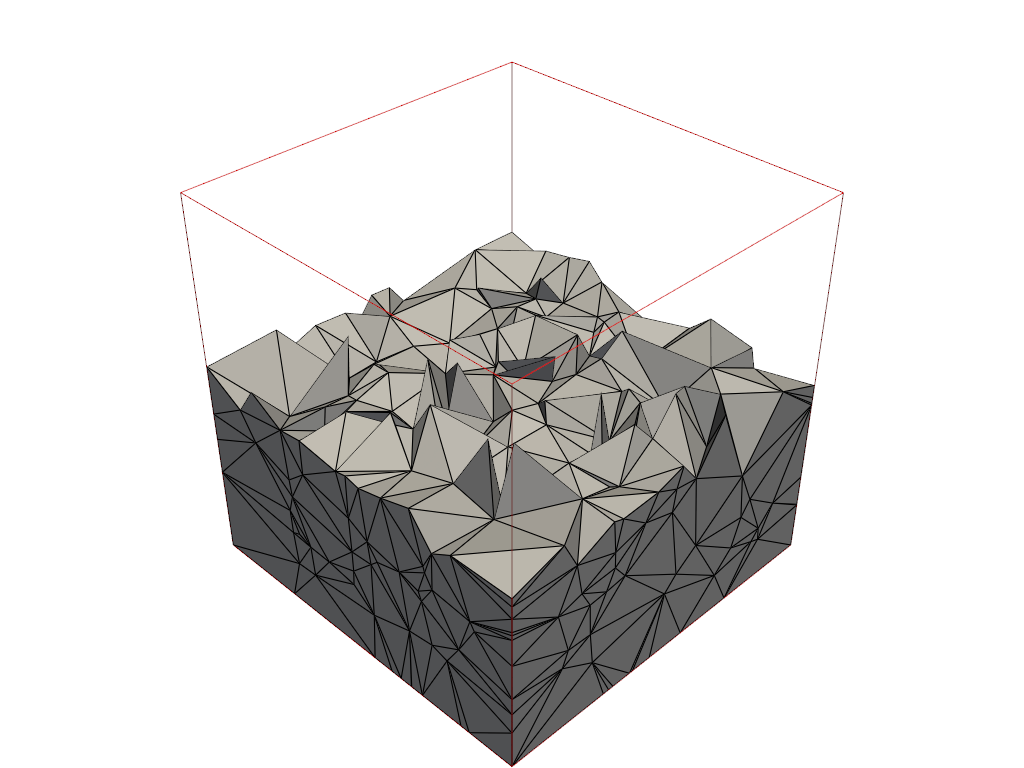}
    \caption{Random mesh}
    \label{fig:randommesh}
    \end{subfigure}
    \begin{subfigure}{0.45\textwidth}
    \centering
    \includegraphics[width=\textwidth]{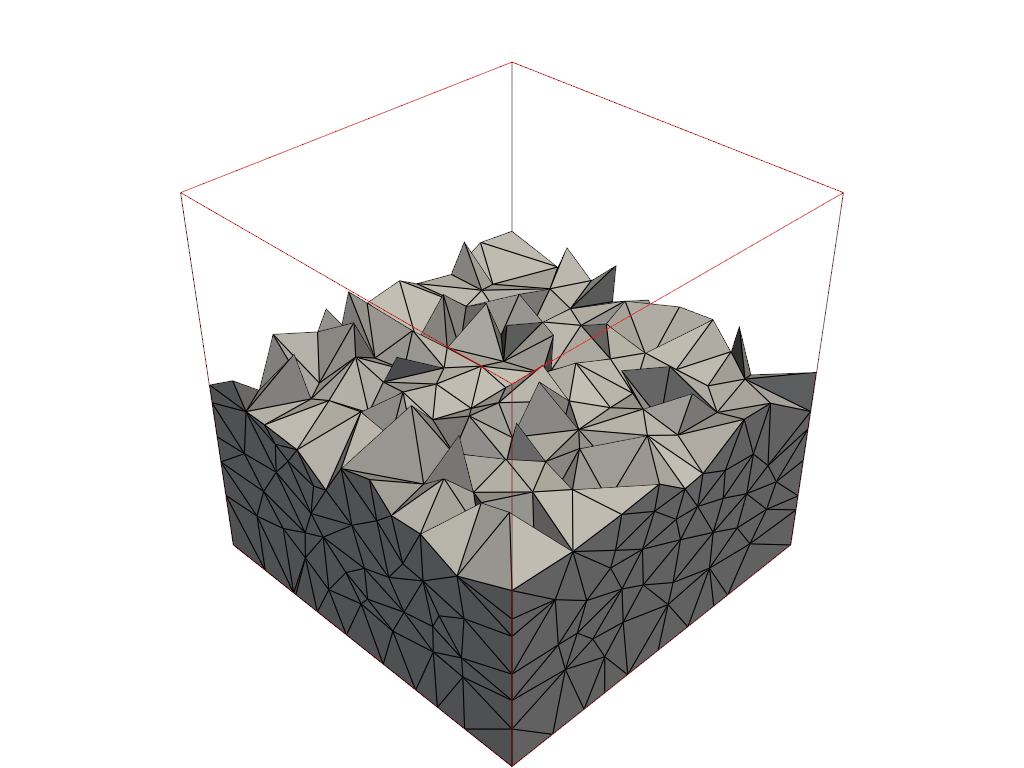}
    \caption{Poisson mesh}
    \label{fig:poissonmesh}
    \end{subfigure}
    \\
    \begin{subfigure}{0.45\textwidth}
    \centering
    \includegraphics[width=\textwidth]{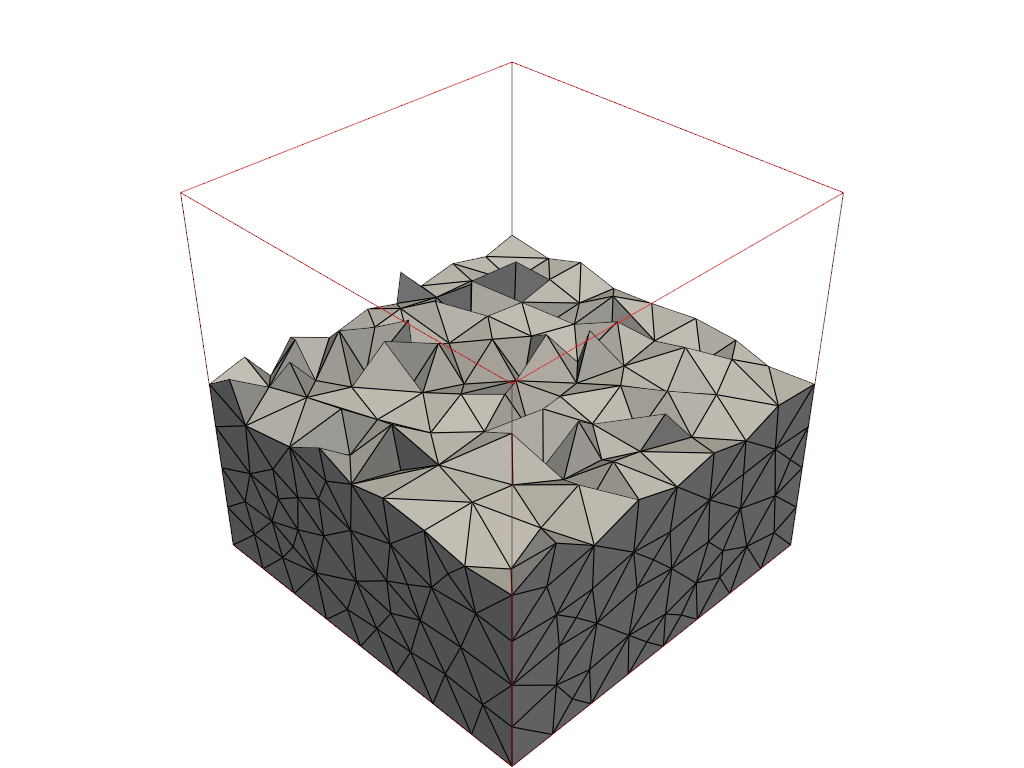}
    \caption{Quality mesh}
    \label{fig:tetgenmesh}
    \end{subfigure}
    \begin{subfigure}{0.45\textwidth}
    \centering
    \includegraphics[width=\textwidth]{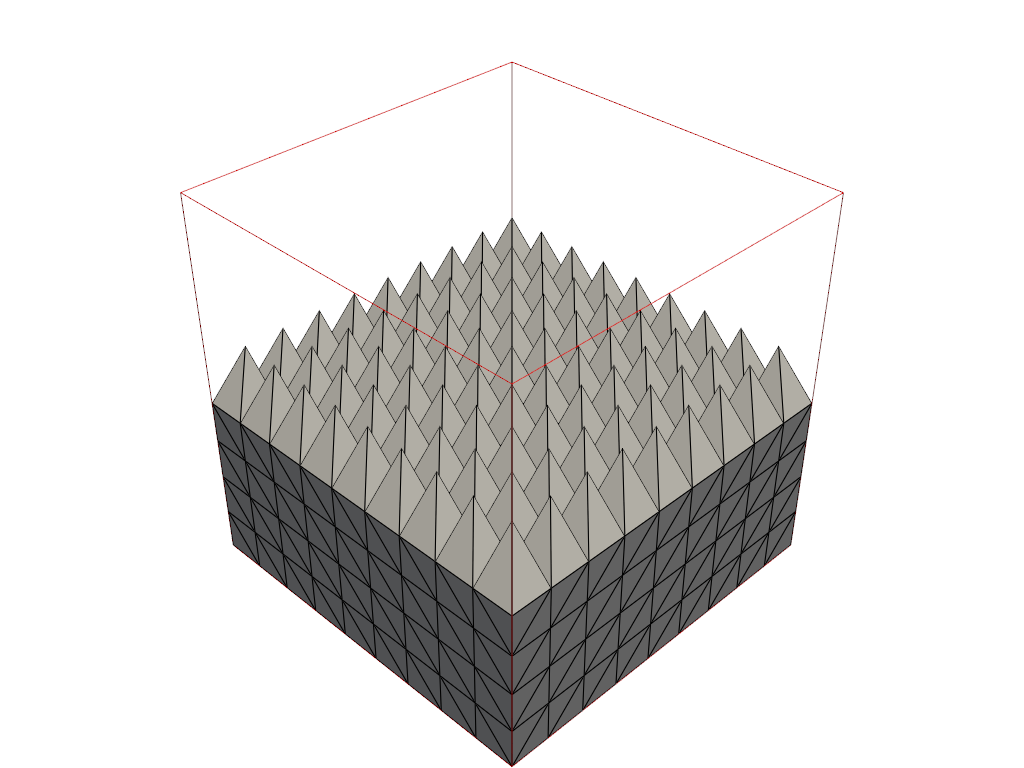}
    \caption{Grid mesh}
    \label{fig:gridmesh}
    \end{subfigure}
    \caption{Example of the meshes generated for the experiments, all meshes have near 5000 vertices. The cubes were cut by a plane to show the interior of the mesh.}
    \label{fig:examplemeshes}
\end{figure*}

\begin{table*}[tb]
    \centering

\resizebox{2\columnwidth}{!}{%
\begin{tabular}{l|lrrr|rrrrrrrr}
    \toprule
    & {} & \multicolumn{2}{c}{Tetrahedral mesh} & \multicolumn{8}{c}{Polyhedral mesh} \\
    \cmidrule(lr){2-5} \cmidrule(lr){6-12}

   {} & \#V &     \#F &     \#T &     \#E &  \#P' & \#B. faces & \#P. w\textquotesingle barriers &  \#PolyTetras & Max tetras &  Avg tetras & \#F' &  Time \\
    \midrule
    \multirow{5}{*}{\begin{turn}{90}Random\end{turn}} 
   & 496   &    5155 &    2462 &    3188 &          748 &            477 &                       182 &           191 &          15 &         3.0 &    3442 &   0.1 \\
   & 989   &   10283 &    4910 &    6361 &         1444 &            885 &                       369 &           344 &          19 &         3.0 &    6817 &   0.3 \\
   & 4997  &   59369 &   29025 &   35340 &         8637 &           5537 &                      2143 &          2153 &          21 &         3.0 &   38989 &   1.7 \\
   & 9988  &  118495 &   57917 &   70565 &        17032 &          11019 &                      4265 &          4164 &          24 &         3.0 &   77617 &   3.3 \\
   & 49999 &  642745 &  318552 &  374191 &        95933 &          62262 &                     24113 &         24203 &          32 &         3.0 &  420174 &  18.9 \\
    \midrule
    \multirow{5}{*}{\begin{turn}{90}Poisson\end{turn}} 
    & 533   &    5196 &    2458 &    3270 &          828 &            352 &                       202 &           168 &          11 &         3.0 &    3566 &   0.1 \\
    & 1002  &    9714 &    4589 &    6126 &         1505 &            710 &                       381 &           355 &          13 &         3.0 &    6630 &   0.3 \\
    & 4995  &   56200 &   27331 &   33863 &         9198 &           4308 &                      2393 &          2280 &          16 &         3.0 &   38069 &   1.8 \\
    & 10557 &  119178 &   58005 &   71729 &        19355 &           9126 &                      4966 &          4606 &          16 &         3.0 &   80532 &   3.5 \\
    & 50149 &  616412 &  304756 &  361804 &       104452 &          49083 &                     28055 &         26313 &          17 &         3.0 &  416146 &  17.2 \\
\midrule
\multirow{5}{*}{\begin{turn}{90}Quality\end{turn}}  
& 513   &    4729 &    2227 &    3014 &          748 &            301 &                       178 &           155 &          12 &         3.0 &    3250 &   0.1 \\
& 996   &    9586 &    4553 &    6028 &         1533 &            641 &                       372 &           360 &          17 &         3.0 &    6568 &   0.3 \\
& 5030  &   53568 &   25986 &   32611 &         8655 &           3413 &                      2210 &          1932 &          15 &         3.0 &   36239 &   1.5 \\
& 10134 &  111654 &   54515 &   67272 &        18088 &           7147 &                      4556 &          3971 &          14 &         3.0 &   75231 &   3.2 \\
& 50027 &  578629 &  285369 &  343286 &        94196 &          37881 &                     24128 &         20154 &          18 &         3.0 &  387486 &  16.3 \\
\midrule
\multirow{5}{*}{\begin{turn}{90}Grid\end{turn}}  
 
& 512   &    4410 &    2058 &    2863 &          690 &             33 &                        33 &             7 &           5 &         3.0 &    3042 &   0.1 \\
& 1000  &    9234 &    4374 &    5859 &         1460 &             53 &                        53 &            14 &           5 &         3.0 &    6320 &   0.3 \\
& 4913  &   50688 &   24576 &   31024 &         8376 &           1741 &                      1741 &           471 &           5 &         3.0 &   34476 &   1.4 \\
& 10648 &  113778 &   55566 &   68859 &        18565 &            721 &                       721 &           224 &           5 &         3.0 &   76771 &   3.4 \\
& 50653 &  567648 &  279936 &  338364 &        93829 &           8998 &                      8998 &          2362 &           5 &         3.0 &  381531 &  15.7 \\
    \bottomrule
\end{tabular}
}
\caption{Table of the experiments  for in-circle criterion. For an input Tetrahedral mesh, ``\#V'' represents the number of vertices, ``\#F'' is the number of faces, ``\#T'' indicates the number of tetrahedrons, and ``\#E' signifies the number of edges. For the output  mesh, ``\#P'' denotes the total number of polyhedrons, ``B. Faces'' refers to the count of barrier-faces, ``P. with barriers'' is the number of polyhedrons containing barrier-faces. ``\#PolyTetras'' represents the number of polyhedrons that are tetrahedrons, ``Max tetras'' is the maximum number of tetrahedrons in a polyhedron, ``Avg tetras'' gives the average number of tetrahedrons per polyhedron, and ``Time'' is the duration in milliseconds that the algorithm takes to execute. The colored numbers are to compare the maximum, in red, and minimum, in blue, value obtained in the experiments of 50k vertices.  }
\label{table:tabla3DAll}
\end{table*}

\begin{table*}[tb]
    \centering

\resizebox{2\columnwidth}{!}{%
\begin{tabular}{l|lrrr|rrrrrrrr}
    \toprule
    & {} & \multicolumn{2}{c}{Tetrahedral mesh} & \multicolumn{8}{c}{Polyhedral mesh} \\
    \cmidrule(lr){2-5} \cmidrule(lr){6-12}

   {} & \#V &     \#F &     \#T &     \#E &  \#P' & \#B. faces & \#P. w\textquotesingle barriers &  \#PolyTetras & Max tetras &  Avg tetras & \#F' &  Time \\
    \midrule
    \multirow{5}{*}{\begin{turn}{90}Random\end{turn}} 
   & 496   &    5155 &    2462 &    3188 &          738 &            238 &                       146 &           158 &          16 &         3.0 &    3431 &   0.0 \\
   & 989   &   10283 &    4910 &    6361 &         1448 &            455 &                       298 &           265 &          26 &         3.0 &    6821 &   0.0 \\
   & 4997  &   59369 &   29025 &   35340 &         8495 &           2777 &                      1836 &          1541 &          18 &         3.0 &   38834 &   0.2 \\
   & 9988  &  118495 &   57917 &   70565 &        16747 &           5428 &                      3576 &          3008 &          22 &         3.0 &   77325 &   0.5 \\
   & 49999 &  642745 &  318552 &  374191 &        92339 &          30704 &                     20275 &         15856 &          26 &         3.0 &  416526 &   2.8 \\
    \midrule
    \multirow{5}{*}{\begin{turn}{90}Poisson\end{turn}} 
    & 533   &    5196 &    2458 &    3270 &          794 &            178 &                       134 &           125 &          12 &         3.0 &    3532 &   0.0 \\
    & 1002  &    9714 &    4589 &    6126 &         1394 &            352 &                       248 &           210 &          14 &         3.0 &    6520 &   0.0 \\
    & 4995  &   56200 &   27331 &   33863 &         8453 &           2246 &                      1654 &          1386 &          16 &         3.0 &   37322 &   0.2 \\
    & 10557 &  119178 &   58005 &   71729 &        18018 &           4853 &                      3496 &          2984 &          23 &         3.0 &   79193 &   0.5 \\
    & 50149 &  616412 &  304756 &  361804 &        95469 &          26172 &                     19457 &         15573 &          18 &         3.0 &  407135 &   2.6 \\
\midrule
\multirow{5}{*}{\begin{turn}{90}Quality\end{turn}}  
& 513   &    4729 &    2227 &    3014 &          711 &            198 &                       137 &           129 &          12 &         3.0 &    3213 &   0.0 \\
& 996   &    9586 &    4553 &    6028 &         1467 &            372 &                       271 &           270 &          13 &         3.0 &    6501 &   0.1 \\
& 5030  &   53568 &   25986 &   32611 &         8234 &           2177 &                      1591 &          1426 &          19 &         3.0 &   35817 &   0.2 \\
& 10134 &  111654 &   54515 &   67272 &        17216 &           4671 &                      3426 &          2957 &          17 &         3.0 &   74360 &   0.5 \\
& 50027 &  578629 &  285369 &  343286 &        89156 &          24895 &                     18064 &         14959 &          19 &         3.0 &  382428 &   3.7 \\
\midrule
\multirow{5}{*}{\begin{turn}{90}Grid\end{turn}}  
 
& 512   &    4410 &    2058 &    2863 &          695 &            311 &                       136 &           190 &          16 &         3.0 &    3048 &   0.0 \\
& 1000  &    9234 &    4374 &    5859 &         1301 &           1086 &                       256 &           298 &          15 &         3.0 &    6161 &   0.0 \\
& 4913  &   50688 &   24576 &   31024 &         7127 &           5386 &                      1615 &          1464 &          20 &         3.0 &   33248 &   0.2 \\
& 10648 &  113778 &   55566 &   68859 &        15944 &          13775 &                      3428 &          3190 &          24 &         3.0 &   74184 &   0.8 \\
& 50653 &  567648 &  279936 &  338364 &        78934 &          68376 &                     17538 &         14850 &          24 &         4.0 &  366755 &   2.6 \\
    \bottomrule
\end{tabular}
}
\caption{Table of the experiments  for area criterion. For an input Tetrahedral mesh, ``\#V'' represents the number of vertices, ``\#F'' is the number of faces, ``\#T'' indicates the number of tetrahedrons, and ``\#E' signifies the number of edges. For the output  mesh, ``\#P'' denotes the total number of polyhedrons, ``B. Faces'' refers to the count of barrier-faces, ``P. with barriers'' is the number of polyhedrons containing barrier-faces. ``\#PolyTetras'' represents the number of polyhedrons that are tetrahedrons, ``Max tetras'' is the maximum number of tetrahedrons in a polyhedron, ``Avg tetras'' gives the average number of tetrahedrons per polyhedron, and ``Time'' is the duration in milliseconds that the algorithm takes to execute. The colored numbers are to compare the maximum, in red, and minimum, in blue, value obtained in the experiments of 50k vertices.}
\label{table:tabla3DAllarea}
\end{table*}


\begin{table*}[!h]
    \centering
    \resizebox{2\columnwidth}{!}{%
    \begin{tabular}{lrrrrrrrrrr}
        \toprule
        {} & \multicolumn{5}{c}{In-circle criterion} & \multicolumn{5}{c}{Area max criterion} \\
        \cmidrule(lr){2-6} \cmidrule(lr){7-11}
        {} &  Random &  Poisson &  Quality &  Grid &  Total  &  Random &  Poisson &  Quality &  Grid &  Total \\
        \cmidrule(lr){1-1} \cmidrule(lr){2-6} \cmidrule(lr){7-11}
        Reduction    &    \textcolor{red}{70.2\%} &     \textcolor{blue}{66.4\%} &         66.6\% &     \textcolor{blue}{66.4\%} &   67.4 &  \textcolor{red}{70.7\%} &     68.8\% &         \textcolor{blue}{68.3\%} &     70.1\% &   69.5\% \\
        Avg Tetras   &     3.0 &      3.0 &          3.0 &      3.0 &    3.0 &   3.0 &      3.0 &          3.0 &      \textcolor{red}{3.2} &    3.0 \\
        Barriers     &    25.0\% &     \textcolor{red}{25.6\%} &         24.9\% &      \textcolor{blue}{8.5\%} &   21.0\% &  \textcolor{red}{21.1\%} &     \textcolor{blue}{18.8\%} &         19.4\% &     \textcolor{red}{21.1} &   20.1\% \\
        Tetrahedrons &    \textcolor{red}{24.8} &     23.5 &         22.0 &      \textcolor{blue}{2.3} &   18.1 &  18.6 &     \textcolor{blue}{16.0} &         17.6 &     \textcolor{red}{21.9} &   18.5 \\
        \bottomrule
        \end{tabular}
    }
        \caption{Summary of the results from Tables~\ref{table:tabla3DAll} and \ref{table:tabla3DAllarea}. The ``Reduction'' row shows the percentage of tetrahedrons removed by the algorithm. The ``Avg tetras'' row indicates the average number of tetrahedrons contained within each polyhedron. The ``Barrier'' row displays the percentage of polyhedrons containing barrier faces. And lastly, the ``Tetrahedrons'' row presents the total number of tetrahedrons remaining in the  mesh. For each row, the colored numbers compare the maximum, in red, and minimum, in blue, value obtained in each distribution experiment.}
        \label{table:ResultasTables}    
\end{table*}


For each kind of mesh, we generate $5$ different sizes, with $500$, $1000$, $5000$, $10000$ and $50000$ vertices. We run the  algorithm with two joining criteria, the in-circle criterion and the Max area criterion, it was made to understand how change the behavior of the quality of the mesh with both criteria. The results are shown in Table~\ref{table:tabla3DAll}, for the in-circle criterion, and the Table~\ref{table:tabla3DAllarea} for the Max area criterion. The tables show the number of vertices, faces, tetrahedrons and edges of the input tetrahedral mesh, the number of polyhedrons, the number of barrier-faces, the number of polyhedrons with barrier-faces, the number of polyhedrons without barrier-faces, the maximum and the average number of tetrahedrons used to create new polygons, and the time in seconds that the algorithm takes to run. 

The summary of the tables can be seen in Table~\ref{table:ResultasTables}. In general, regardless of the point distribution of the meshes, the algorithm merges, on average, 3 tetrahedron per polygon, reducing the tetrahedron mesh by around $70\%$ of the elements, and about one-fifth of the polyhedra contain barrier faces. This is even true for points not in general position, as it is the case of Grid meshes.

We can see differences between the two joining criterion, the in-circle criterion tends to join more tetrahedrons than the area max criterion, and the generate polyhedrons will result in less barrier-edge tips, but in they generated meshes will contains in average more tetrahedrons.

The only statistic that seems to vary depending on the distribution is the percentage of polyhedrons that remain tetrahedrons after the algorithm was applied. Poisson's meshes have the lowest percentage of tetrahedrons, while Grid meshes have the highest. However, on average, all meshes retain approximately $18.6\%$ of their elements as tetrahedrons.  

\subsection{Comparison with Voronoi}

In order to do a simple comparison between  polyhedral meshes generated by our algorithm and  Voronoi meshes, we compute the number of vertices, edges, faces and polyhedra of the dual diagram of the quality tetrahedral meshes used as input shown in Table~\ref{table:tabla3DAllarea}.  Table~\ref{table:Voronoi} shows at the left the number of vertices, edges, faces and tetrahedra of the quality tetrahedral mesh, at the center, the same information for the Voronoi mesh and at the right for the polyhedral mesh generated by our algorithm. It can be observed that the number of vertices of the polyhedral mesh is less than in the Voronoi mesh, the number of Voronoi faces is slightly smaller than in the polyhedral mesh and the number of Voronoi cells is less than the polyhedral cells. It is worth to mention that this Voronoi meshes are not constrained to the boundary of the domain. Constrained Voronoi meshes require additional  vertices and faces to fit the domain geometry.

 \begin{figure}[h]
  \centering
    \includegraphics[width=0.9\linewidth]{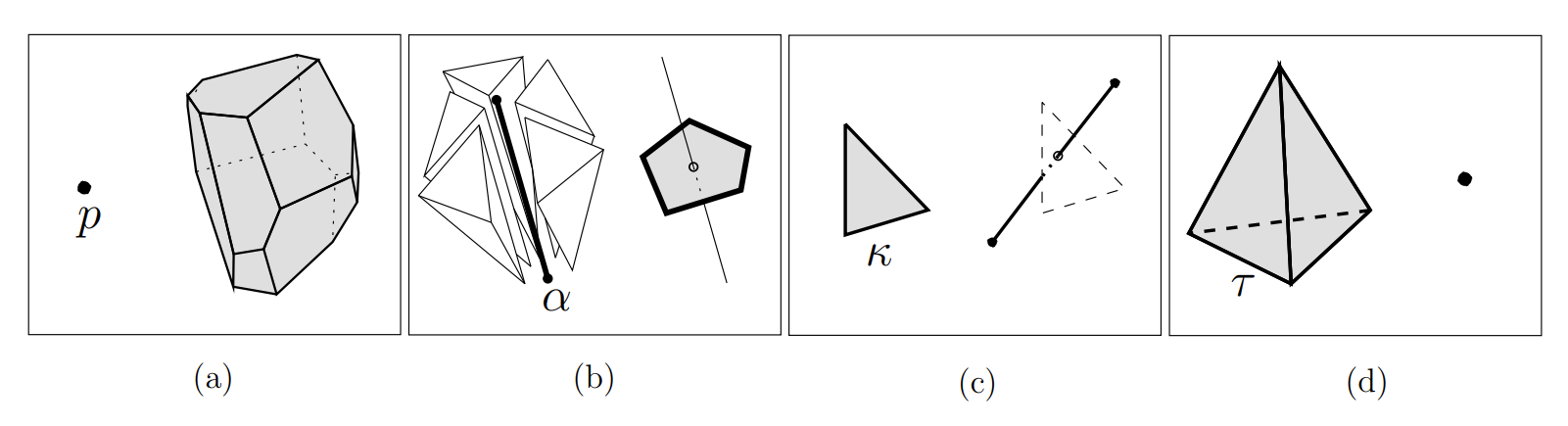}
    \caption{Duality of the 3D Delaunay tetrahedralization $DT$ to Voronoi diagram $VD$, in (a) each point $p$ of $DT$ is enclosed by a Voronoi cell, in (b) each edge $DT$ has associated a Voronoi face, in (c) each triangle face of $DT$ has associated a Voronoi edge of $VD$, and in (d) The circumcenter of each tetrahedron circumsphere is a Voronoi point of $VD$. Source~\cite{3DDelaunayToVoronoi}.}
    \label{fig:3DDelaunayToVoro}
  \end{figure}

\begin{table*}[htb]
    \centering
    \resizebox{1.8\columnwidth}{!}{%
    \begin{tabular}{rrrr|rrr|rrr}
        \toprule
        \multicolumn{4}{c}{ Delaunay tetrahedral mesh} & \multicolumn{3}{c}{Voronoi Mesh} & \multicolumn{3}{c}{Our Polyhedral Mesh} \\
        \cmidrule(r){1-4} \cmidrule(r){5-7} \cmidrule{8-10}
        Vertices & Faces & Tetrahedrons & Edges & Voronoi Cells & Voronoi Faces & Vertices & Vertices & Polyhedra & Faces \\
        \midrule
513   &  4729  &  2227  &  3014  &  513   &  3014  &  2227  & 513   &          748 &      3250 \\
996   &  9586  &  4553  &  6028  &  996   &  6028  &  4553  & 996   &         1533 &      6568 \\
5030  & 53568  & 25986  & 32611  & 5030   & 32611  & 25986  & 5030  &         8655 &     36239 \\
10134  &111654  & 54515  & 67272  &10134   & 67272  & 54515 & 10134 &        18088 &     75231  \\
50027  &578629  &285369  &343286  &50027   &343286  &285369 & 50027 &        94196 &    387486  \\
        \bottomrule
    \end{tabular}
    }
    \caption{Comparison between the Voronoi mesh and the quality mesh using the in-circle criterion. The table shows the number of vertices, edges, polygons, and faces for each mesh type.}
    \label{table:Voronoi}
\end{table*}

\section{Conclusions and future work}\label{sec:conclsion}

In this paper, we have presented the theoretical concepts and a preliminary version of an algorithm  designed to convert a tetrahedral mesh into a polyhedral mesh. We have introduced the concept of terminal/face regions  in order to guide the polyhedral construction. 
 The preliminary results show that the algorithm reduces the number of polyhedrons by approximately $70\%$, and that the average number of tetrahedra per polyhedron is $3$. The algorithm also retains approximately $20\%$ of the tetrahedra in the polyhedral mesh. Those results seems to be almost the same, independent of the  chosen joining criterion. This means that the concept of terminal face region as defined in this paper needs to be improved by considering joining strategies that also considers quality criteria of the generated polyhedra. Our ongoing work is taken in consideration the last findings. Moreover we have to test if these polyhedra are useful in the context of VEM simulations.

\bibliographystyle{siam}
\bibliography{ltexpprt_doublecolumn}

\end{document}